\documentstyle[12pt]{article}
\begin{document}

\baselineskip=21pt

\centerline{\bf Teleportation of Entangled States of a Vacuum-One
Photon Qubit} \vskip6mm

\centerline{Egilberto Lombardi$^{1}$, Fabio Sciarrino$^{1}$, Sandu
Popescu$^{2,3}$ } \centerline{ and Francesco De Martini$^{1}$}

\vskip 6mm

\centerline{\it ${}^{(1)}${Istituto Nazionale di Fisica della
Materia, Dipartimento di Fisica, }} \centerline{\it{ Universita'
``La Sapienza'', Roma, 00185 Italy}}

\vskip 3mm

\centerline{\it ${}^{(2)}$ {H. H. Wills Physics Laboratory,
University of Bristol, }} \centerline{\it{Tyndall Avenue, Bristol
BS8-1TL, U.K.}}

\vskip 3mm

\centerline{\it ${}^{(3)}${BRIMS, Hewlett-Packard Labs., Stoke
Gifford, Bristol BS12-6QZ, U.K.}}

\vskip6mm

We report the experimental realization of teleporting an entangled
qubit. The qubit is physically implemented by a two-dimensional
subspace of states of a mode of the electromagnetic field,
specifically, the space spanned by the vacuum and the one photon
state. Our experiment follows along lines suggested by H. W. Lee
and J. Kim, Phys. Rev. A, 63, 012305 (2000) and E.Knill,
R.Laflamme and G.Milburn Nature 409: 46 (2001).


\vskip10mm

In their pioneering paper C.H. Bennett, G.Brassard, C. Crepeau, R.
Jozsa, A. Peres and W. Wootters introduced the concept of
teleportation of a quantum state \cite{1}. Since then
teleportation has come to be recognized as one of the basic
methods of quantum communication and, more generally, as one of
the basic ideas of the whole field of quantum information.
Following the original teleportation paper and its
continuous-variables version \cite{2} an intensive experimental
effort started for the practical realization of teleportation.
Quantum state teleportation (QST) has been realized in a number of
experiments \cite{3, 4, 5, 6}. In a beautiful example of
ingenuity, although starting from a common theme, each of these
experiments followed a completely different route and principle.
In the present paper we report a new teleportation experiment
following yet another route. In our experiment we consider a qubit
which is physically realized not by a particle but by a {\it mode}
of the electromagnetic (e.m.) field, and whose orthogonal basis
states $\left| 0\right\rangle $, $\left| 1\right\rangle $ are the
vacuum state and the one-photon state respectively. Furthermore,
we teleport entangled states of this qubit. We designed our scheme
by adapting a method proposed by Knill, Laflamme and Milburn
\cite{7} to make it experimentally easily feasible. We later
learned that our method is identical to that proposed by H.W.Lee
and J.Kim \cite{8} and also closely related to \cite{9}. In our
experiment the role of the two particles in a singlet state which
constitute the non-local communication channel in the original
teleportation scheme \cite{1} is played by a photon in an equal
superposition of being at Alice and Bob
$|\Psi\rangle = 2^{-%
%TCIMACRO{\UNICODE[m]{0xbd}}%
%BeginExpansion
{\frac12}%
%EndExpansion
}(\left| Alice\right\rangle +\left| Bob\right\rangle )$ where $\left|
Alice\right\rangle $ and $\left| Bob\right\rangle $ represent the photon
located at Alice and Bob respectively. The scheme seems puzzling. Indeed
entanglement is considered the basis of teleportation and here we don't even
have two particles, let alone two particles in an entangled state. The
puzzle is solved however by noting that in second quantization the state of
the non-local channel reads as $\left| \Phi \right\rangle _{singlet}$= $2^{-%
%TCIMACRO{\UNICODE[m]{0xbd}}%
%BeginExpansion
{\frac12}%
%EndExpansion
}(\left| 1\right\rangle _{A}\left| 0\right\rangle _{B}-\left|
0\right\rangle _{A}\left| 1\right\rangle _{B})$ where the labels
$A$ and $B$ represent two different modes of the e.m. field, with
wavevectors (wv) $k_{A}$ and $k_{B}$ one directed towards Alice
and the other towards Bob. The\ mode indices 0 and 1 denote the
Fock state population by zero (vacuum) and one photon
respectively. In effect the role of the two entangled quantum
systems which form the non-local channel are played by the e.m.
fields of Alice and Bob. In other words the {\it field's modes
}rather than the photons associated with them should be properly
taken as the information and entanglement carriers, i.e. {\it
qubits}. (In the context of Bell's inequalities,the nonlocal
aspects of a single photon have been discussed in
\cite{10},\cite{11},\cite{12} and \cite{13}.)

Of course, in order to make use of the entanglement present in
this picture we need to use the second quantization procedure of
creation and annihilation of particles and/or use states which are
superpositions of states with different numbers of particles.
Another puzzling aspect of this second quantized picture is the
need to define and measure the relative phase between states with
different number of photons, such as the relative phase between
the vacuum and one photon state in Eq. 1 below. That we can
associate a relative phase between the {\it vacuum }and anything
else seems most surprising, but it is less so if we recall the
more familiar case of a coherent state, where the relative phase
between the different photon number states in the superposition is
reflected physically in the phase of the classical electric field.
To be able to control these relative phases we need, by analogy
with classical computers, to supply all gates and all
sender/receiving stations of a quantum information network with a
common {\it clock} signal, e.g. provided by an ancillary photon or
by a multi-photon, Fourier transformed coherent e.m. pulse
\cite{14}. These concepts will be fully demonstrated by the
present experiment.

The quantum system whose state we want to teleport is physically represented
by another mode of the e.m. field, one with wv $k_{S}$. Again we consider
only a two dimensional Hilbert space of this mode, i.e. spanned by $\left|
0\right\rangle _{S}$ and $\left| 1\right\rangle _{S}$. Thus the {\it mode} $%
k_{S}$ can be considered the {\it qubit }to be{\it \ }teleported. Suppose
now that the qubit $k_{S}$ is in an arbitrary {\it pure} state
\begin{equation}
\alpha \left| 0\right\rangle _{S}+\beta \left| 1\right\rangle _{S}
\end{equation}

The overall state of the system and the non-local channel is then:
\begin{eqnarray}
\left| \Phi _{total}\right\rangle &=&2^{-%
%TCIMACRO{\UNICODE[m]{0xbd}}%
%BeginExpansion
{\frac12}%
%EndExpansion
}(\alpha \left| 0\right\rangle _{S}+\beta \left| 1\right\rangle _{S})(\left|
1\right\rangle _{A}\left| 0\right\rangle _{B}-\left| 0\right\rangle
_{A}\left| 1\right\rangle _{B})  \nonumber \\
&=&2^{-%
%TCIMACRO{\UNICODE[m]{0xbd}}%
%BeginExpansion
{\frac12}%
%EndExpansion
}\alpha \left| \Psi ^{1}\right\rangle _{SA}\left| 1\right\rangle _{B}+2^{-%
%TCIMACRO{\UNICODE[m]{0xbd}}%
%BeginExpansion
{\frac12}%
%EndExpansion
}\beta \left| \Psi ^{2}\right\rangle _{SA}\left| 0\right\rangle
_{B} + \nonumber \\ &&+\frac{1}{2}\left| \Psi ^{3}\right\rangle
_{SA}(\alpha \left|
0\right\rangle _{B}+\beta \left| 1\right\rangle _{B}) + \frac{1}{2}%
\left| \Psi ^{4}\right\rangle _{SA}(\alpha \left| 0\right\rangle _{B}-\beta
\left| 1\right\rangle _{B})
\end{eqnarray}
where the states $\left| \Psi ^{j}\right\rangle _{SA}$,
$j=1,2,3,4$ are defined below in Eq. 3. The teleportation proceeds
with Alice performing a partial Bell measurement. She combines the
modes\ $k_{S}$ and $k_{A}$ on a symmetric (i.e. 50:50) beam
splitter $BS_{A}$ whose output modes $k_{1}$ and $k_{2}$ are
coupled to two detectors $D_{1}$ and $D_{2}$, respectively (see
Figure 1). The action of $BS_{A}$ on the field operators is
expressed by:
$\widehat{a}^{\dagger}_{S}=
2^{-%
%TCIMACRO{\UNICODE[m]{0xbd}}%
%BeginExpansion
{\frac12}%
%EndExpansion
}(\widehat{a}^{\dagger}_{1}+\widehat{a}^{\dagger}_{2})$%
; $\widehat{a}^{\dagger}_{A}=2^{-%
%TCIMACRO{\UNICODE[m]{0xbd}}%
%BeginExpansion
{\frac12}%
%EndExpansion
}(\widehat{a}^{\dagger}_{1}-\widehat{a}^{\dagger}_{2})$
where
labels $1,$ $2$ refer to modes $k_{1}$, $k_{2}$. As a consequence
we obtain:
\begin{eqnarray}
\left| \Psi ^{1}\right\rangle _{SA} &=&\left| 0\right\rangle
_{S}\left| 0\right\rangle _{A} =\left| 0\right\rangle _{1}\left|
0\right\rangle _{2},\nonumber \\ \left| \Psi ^{2}\right\rangle
_{SA}&=&\left|
1\right\rangle _{S}\left| 1\right\rangle _{A} = 2^{-%
%TCIMACRO{\UNICODE[m]{0xbd}}%
%BeginExpansion
{\frac12}%
%EndExpansion
}(\left| 2\right\rangle _{1}\left| 0\right\rangle _{2} +\left|
0\right\rangle _{1}\left| 2\right\rangle _{2}),  \nonumber \\
\left| \Psi ^{3}\right\rangle _{SA} &=&2^{-%
%TCIMACRO{\UNICODE[m]{0xbd}}%
%BeginExpansion
{\frac12}%
%EndExpansion
}(\left| 0\right\rangle _{S}\left| 1\right\rangle _{A}-\left| 1\right\rangle
_{S}\left| 0\right\rangle _{A})=\left| 1\right\rangle _{1}\left|
0\right\rangle _{2},  \nonumber \\
\left| \Psi ^{4}\right\rangle _{SA} &=&2^{-%
%TCIMACRO{\UNICODE[m]{0xbd}}%
%BeginExpansion
{\frac12}%
%EndExpansion
}(\left| 0\right\rangle _{S}\left| 1\right\rangle _{A}+\left| 1\right\rangle
_{S}\left| 0\right\rangle _{A})=\left| 0\right\rangle _{1}\left|
1\right\rangle _{2}
\end{eqnarray}
The state $\left| \Psi ^{3}\right\rangle _{SA}$ is a Bell type
state \cite{1}. From Eq. 3 we see that $\left| \Psi
^{3}\right\rangle _{SA}$ leads to a single photon arriving at the
detector $D_{1}$ and no photons at $D_{2}$. Similarly, $\left|
\Psi ^{4}\right\rangle _{SA}$ is a Bell type state and it leads to
a single photon arriving at the detector $D_{2}$ and no photons at
$D_{1}$. In both these cases the teleportation is successful.
Indeed, when Alice finds $\left| \Psi ^{3}\right\rangle _{SA}$
Bob's e.m. field ends up in the state $\left| \Phi \right\rangle $
= $(\alpha \left| 0\right\rangle _{B}+\beta \left| 1\right\rangle
_{B})$ which is identical to the state to be teleported, while
when Alice finds $\left| \Psi ^{4}\right\rangle _{SA}$, Bob ends
up with the state $\left| \Phi _{\pi }\right\rangle $ = $(\alpha
\left| 0\right\rangle _{B}-\beta \left| 1\right\rangle _{B})$=
$\sigma _{z}\left| \Phi \right\rangle $ which is identical to the
state to be teleported up to a phase shift $\Delta $ = $\pm \pi $.
The states $\left| \Phi \right\rangle $ and $\left| \Phi _{\pi
}\right\rangle $ are connected by a unitary transformation
expressed by the Pauli spin operator $\sigma _{z}$. Bob can easily
correct the phase shift $\Delta $ upon finding out Alice's result
\ In practice this phase correction procedure, generally referred
to as ``active teleportation'' \cite{3} is carried out
automatically by means of a fast electro-optic Pockels cell (EOP)\
inserted in mode $k_{B}$ and triggered by $D_{2}$. On the other
hand, when Alice finds $\left| \Psi ^{1}\right\rangle _{SA}$ or
$\left| \Psi ^{2}\right\rangle _{SA}$ the teleportation fails.
From Eq 3 we see that teleportation is successful in 50\% of the
cases.

A major technical difficulty in the above teleportation scheme is
the preparation and manipulation of the pure states to be
teleported. Indeed, they are superpositions of the vacuum and
one-photon states of the mode $k_{S}$. Manipulating such states
and, in particular having control about the relative phase between
the vacuum and one-photon states is quite problematic. This can be
realized in principle, for example by homodyning techniques as
described in \cite{15}. Here however, we avoid the problem
altogether, by teleporting appropriate {\it entangled} states
instead of pure ones. The states we consider are of the form
\begin{equation}
\left| \Psi \right\rangle _{S\widetilde{a}}=(\alpha \left| 0\right\rangle
_{S}\left| 1\right\rangle _{\widetilde{a}}+\beta \left| 1\right\rangle
_{S}\left| 0\right\rangle _{\widetilde{a}})
\end{equation}
where $k_{\widetilde{a}}$ is an ``ancilla'' mode. These states are
in fact simple single-photon states and can be easily obtained by,
say, letting a single photon impinge on a beam-splitter\ ($BS_{S}$
in Fig. 1) with reflectivity $r_{S}$ and transmissivity $t_{S}$,
$k_{\widetilde{a}}$ being the reflected mode and $k_{S}$ the
transmitted one. For the sake of
simplicity and without loss of generality we assume that $\alpha $ and $%
\beta $ are $real$ numbers.

To summarize, in our experiment we have four qubits: $k_{A}$ and $k_{B}$
which constitute the non-local communication channel,\ $k_{S}$ which
represents the system, i.e. the qubit to be teleported and $k_{\widetilde{a}%
} $ the ancilla. The special states of these four qubits which are used in
the experiment are physically implemented by exactly two photons. The state
of the qubit $k_{S}$ is teleported to Bob into the state of the qubit $k_{B}$%
, thus the overall state $\left| \Psi \right\rangle _{S\widetilde{a}}$ will
now be transferred into the state of the qubits $k_{B}$ and $k_{\widetilde{a}%
}$. To verify that the state has been teleported we transmit the qubit $k_{%
\widetilde{a}}$ to Bob. The QST\ verification consists simply by
mixing the modes $k_{B}$ and $k_{\widetilde{a}}$ at a
beam-splitter $(BS_{B})$ similar to the one which was used to
produce the state to be teleported $\left| \Psi \right\rangle
_{S\widetilde{a}}$. We shall see that the optimum QST\
verification, viz implying the maximum {\it visibility} $V$ of the
corresponding interferometric patterns is obtained by adopting
equal optical parameters for both $BS_{S}$ and $BS_{B}$,
i.e.$\;\left| r_{S}\right| =\left| r_{B}\right| =\alpha $ and
$\left| t_{S}\right| =\left| t_{B}\right| =\beta $. This
verification procedure is generally referred to as ``passive
teleportation'' \cite{3}. Finally, note that the ``ancillary''
single photon emitted on mode $k_{\widetilde{a}}$ indeed provides
the ``clock'' pulse that is needed to retrieve at Bob's side the
{\it full information} content of the vacuum state $\left|
0\right\rangle _{B}$ entangled within the nonlocal teleportation
channel,\ i.e. with the {\it singlet} state $\left| \Phi
\right\rangle _{singlet}$ \cite{14}.

The experimental set-up is shown in Fig. 1. A nonlinear $LiO_{3}$ crystal
slab, 1.5 mm thick with parallel anti-reflection coated faces, cut for Type
I phase-matching is pumped by a single mode UV cw argon laser with
wavelength (wl) $\lambda _{p}=$ $363.8nm$ and with an average power $\simeq
100mW$ . The UV laser beam was focused close to the crystal by a lens with
focal length = 2m in order to maximize the collection efficiency by the
Alice's detector system of the spontaneous parametric down-conversion (SPDC)
fluorescence \cite{16}. The two SPDC emitted photons have equal wl $\lambda
=727,6nm$ and are spatially selected by two pinholes with equal apertures
with diameter $0.5mm$ placed at a distance of $50cm$ from the crystal. One
of the photons generates on the two output modes $k_{A}$ and $k_{B}$ of a
50:50 beam splitter $(BS)$ the singled state $\left| \Phi \right\rangle
_{singlet}$ providing the nonlocal teleportation channel. The other photon\
generates the state $\left| \Psi \right\rangle _{S\widetilde{a}}$ i.e. the
quantum superposition of the state to be teleported and the one of the
ancilla at the output of a {\it variable} beam-splitter\ $BS_{S}$ consisting
of the combination of a $\lambda /2$ polarization rotator and of a calcite
crystal. Furthermore micrometric changes of the mutual phase $\varphi $ of
the $k_{S}$ and $k_{A}$ modes interfering on $BS_{A}$ were obtained by a
piezoelectrically driven mirror $M$. All detectors were Si-avalanche
EG\&G-SPCM200 counting modules having nearly equal quantum efficiencies $%
QE\approx $ $0.45$. Before detection the beams were IF\ filtered
within a bandwidth 20nm. In Figure 1 the complete scheme for
``active'' teleportation is shown, including the high-voltage
Pockels cell (EOP) inserted on the mode $k_{B}$. In the same
figure is reported the interferometric scheme for ``passive
teleportation'' which is also adopted for the verification of the
correct implementation of the ``active'' protocol, as we shall
see.

We have realized experimentally the {\it passive} teleportation protocol. By
this we mean that Bob does not modify his state according to the results
obtained by Alice. Instead Bob passes his state unmodified to the
verification stage. The verification stage consists in combining the mode $%
k_{s}$ (which now contains the teleported state) with the ancilla mode $k_{%
\widetilde{a}}$ at a beam-splitter $BS_{B}$, as said. In order to check the
overall mode alignment we first checked at Alice's site the 2-photon
Ou-Mandel interference across the beam-splitter $BS_{A}$ between the modes $%
k_{S}$ and $k_{A}$ that are coupled to detectors $D_{1}$ and $D_{2}$
respectively. We obtained a 2-photon interference pattern with a visibility $%
V_{A}\approx 0.96$. In a similar way we checked, at Bob's site the Ou-Mandel
interference across $BS_{B}$ between the modes $k_{B}$ and $k_{\widetilde{a}%
} $ coupled to the respective detectors $D_{1}^{\ast }$, $D_{2}^{\ast }$
obtaining: $V_{B}\approx 0.92$.\ The QST verification experiment has been
carried out first with a 50:50 beam splitter $BS_{S}$, i.e. with optical
parameters: $\left| r_{S}\right| =\left| t_{S}\right| =2^{-%
%TCIMACRO{\UNICODE[m]{0xbd}}%
%BeginExpansion
{\frac12}%
%EndExpansion
}$. The maximum visibility of the verification fringe pattern is obtained by
selecting the same values of the parameters for the test beam splitter $%
BS_{B}$, as said. Then we measured the coincidence counts between $D_{1}$, $%
D_{2}$ and $D_{1}^{\ast }$, $D_{2}^{\ast }$. By a straightforward
calculation we expect
\begin{equation}
(D_{1}-D_{1}^{\ast })=(D_{2}-D_{2}^{\ast })=%
%TCIMACRO{\UNICODE[m]{0xbd}}%
%BeginExpansion
{\frac12}%
%EndExpansion
\sin ^{2}\frac{\varphi }{2},(D_{1}-D_{2}^{\ast })=(D_{2}-D_{1}^{\ast })=%
%TCIMACRO{\UNICODE[m]{0xbd}}%
%BeginExpansion
{\frac12}%
%EndExpansion
\cos ^{2}\frac{\varphi }{2}
\end{equation}
where $(D_{i}-D_{j}^{\ast })$ expresses the probability of a coincidence
detected by the pair $D_{i}$, $D_{j}^{\ast }$ in correspondence with the
realization of either one of the states: $\left| \Psi ^{3}\right\rangle
_{SA} $, $\left| \Psi ^{4}\right\rangle _{SA}$. The experimental plots shown
in Figure 2, obtained by varying the position $X=$ $(2)^{-3/2}\lambda
\varphi /\pi $ of the mirror M, are in agreement with the theory, Eq. 5.
This agreement is further substantiated by the data reported in Fig. 3
corresponding to a similar QST\ verification experiment carried out with a
different set of optical parameters for $BS_{B}$: $\left| r_{B}\right|
^{2}=0.20$, $\left| t_{B}\right| ^{2}=$ $0.80$. Precisely, each experimental
point of Fig. 3 corresponds to an experiment equal to the one referred to by
Fig. 2. The {\it visibility} of each sinusoidal fringe pattern is then
reported. Note that the maximum visibility $V$ is attained for values of $%
\alpha ^{2}=1-\beta ^{2}$ that are equal to $\left| r_{B}\right| ^{2}$ or to
$\left| t_{B}\right| ^{2}$ depending on which pair of detectors are excited.
The two different peaks collapse into only one maximum with theoretical $V=1$
in the fully symmetric case: $\left| r_{B}\right| ^{2}=\left| t_{B}\right|
^{2}=$ $\alpha ^{2}=\beta ^{2}=%
%TCIMACRO{\UNICODE[m]{0xbd}}%
%BeginExpansion
{\frac12}%
%EndExpansion
$. In Fig. 3 is also reported a single experimental value $V\simeq 0.91$
related to the symmetric case.

Note that by assuming perfect detectors, i.e. with $QE=1$, the
above QST verification procedure involving the ancilla mode
$k_{\widetilde{a}}$ enables a fully {\it noise-free} teleportation
procedure. Indeed, if no photons are detected at Alice'site, i.e.
by $D_{1}$and/or $D_{2}$, while photons are detected at Bob's site
by $D_{1}^{\ast }$ and/or $D_{2}^{\ast }$ we can safely conclude
that the ``idle'' Bell state $\left| \Psi ^{1}\right\rangle _{SA}$
has been created. If, on the other hand, no photons are detected
at Bob's site while photons are detected at Alice's site, we must
conclude that the other ``idle'' Bell state $\left| \Psi
^{2}\right\rangle _{SA}$ has been realized. The data collected in
correspondence with these ``idle'' events can automatically\ be
discarded by the electronic coincidence circuit. In addition to
that, note that the effect of \ the above verification procedure
involving the ancilla mode $k_{\widetilde{a}}$ keeps holding
within the {\it active} teleportation scheme. Indeed, if the
$D_{2}-driven$ EO phase-modulator works correctly within the {\it
active} scheme, the detector $D_{2}^{\ast }$ should be found to be
{\it always} inactive.

Our present effort is directed towards the completion of the
teleportation picture by the realization of  the ``active''
scheme. The main technical problem resides in the relatively large
time needed to activate a high-voltage EO device by a single
photon detection. The best result we have attained so far for the
1KV switching time across an\ EO modulator is about 10 nsec. This
figure would enable us to achieve the goal in the near future by
the adoption of small $\lambda /2-voltage$ EO devices possibly in
conjunction with the use of optical fibers. We are greatly
indebted with the FET European Network on Quantum Information and
Communication (Contract IST-2000-29681:ATESIT) and with M.U.R.S.T.
for funding.

\centerline{\bf Figure Captions}

\vskip 8mm

\parindent=0pt

\parskip=3mm

1. Experimental apparatus realizing the ``active'' and ``passive''
quantum state teleportation (QST).

2. Interferometric fringe patterns obtained by coincidence experiments
involving different pairs of detectors within a {\it passive} QST
verification procedure.

3. Visibility $V$ of the coincidence fringe patterns Vs the superposition
parameter $\alpha ^{2}=1-\beta ^{2}$ obtained by two different pairs of
detectors within a {\it passive} QST\ experiment and for an unbalanced beam
splitter $BS_{B}$ with optical parameters $\left| r_{B}\right| ^{2}=0.20$, $%
\left| t_{B}\right| ^{2}=$ $0.80$. The continuous lines represent the
corresponding theorical expectations. A single experimental value of $V$ for
the fully symmetric case $\left| r_{B}\right| ^{2}=\left| t_{B}\right| ^{2}=$
$\alpha ^{2}=0.50$ is also reported.

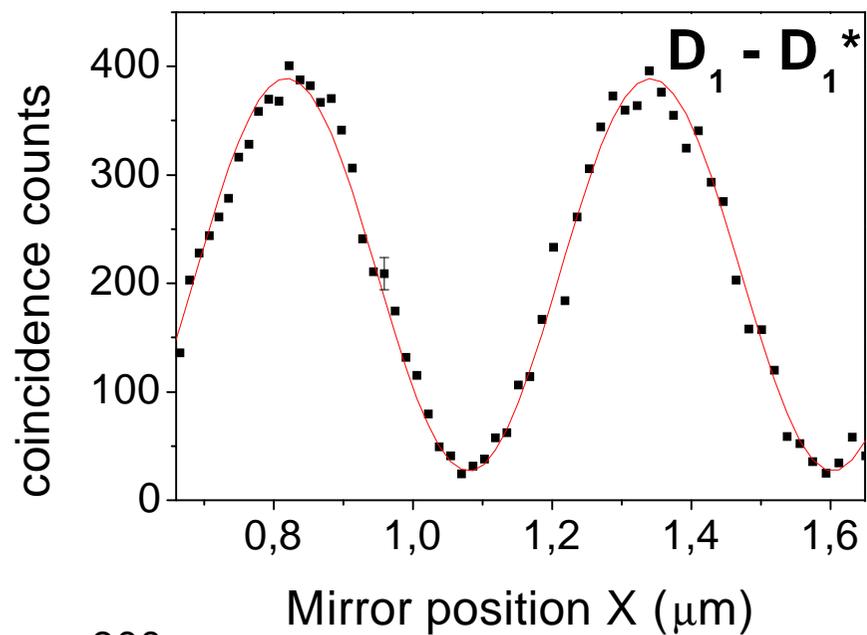
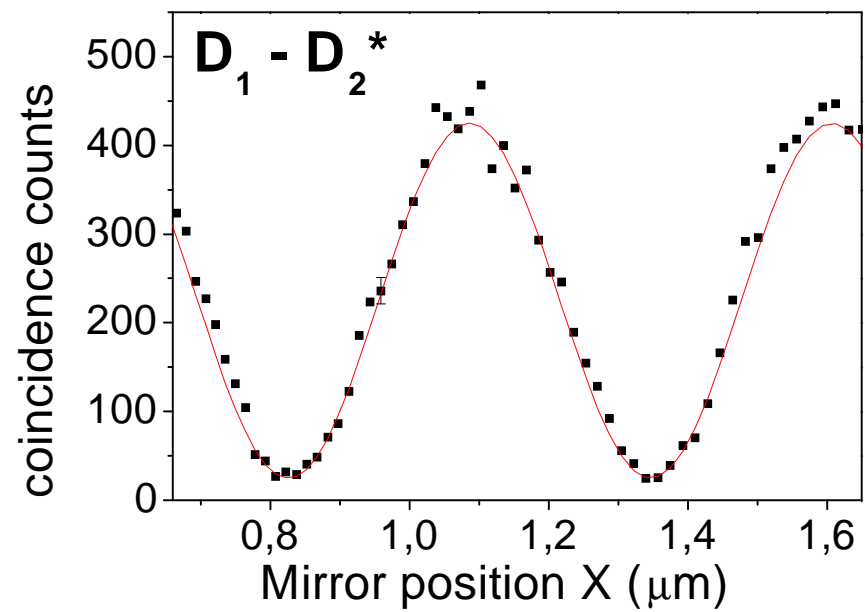
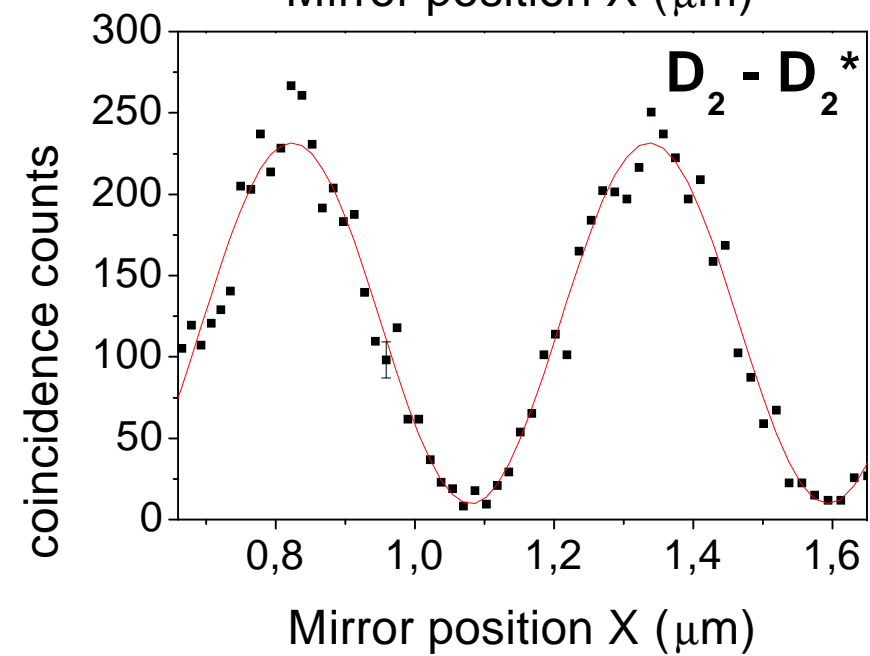
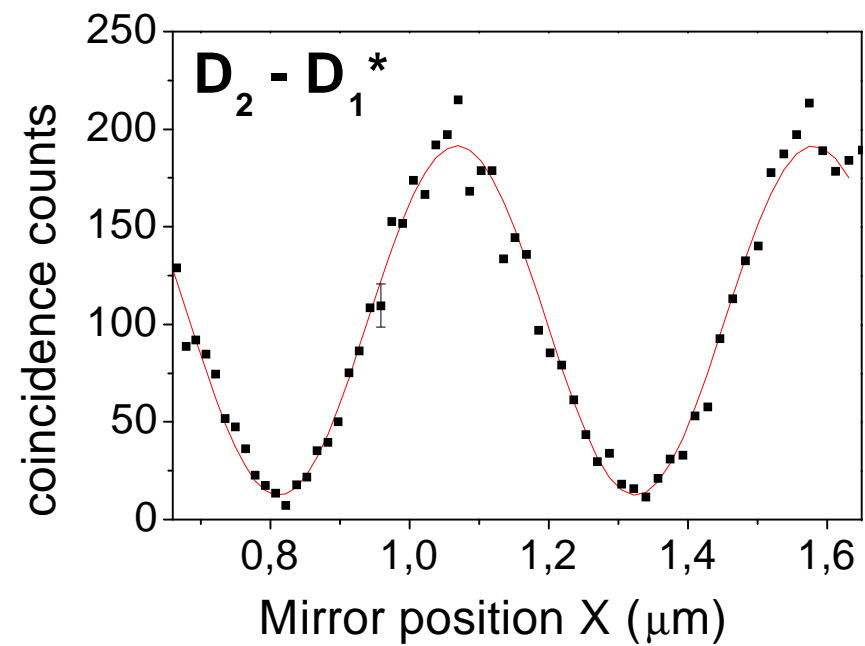
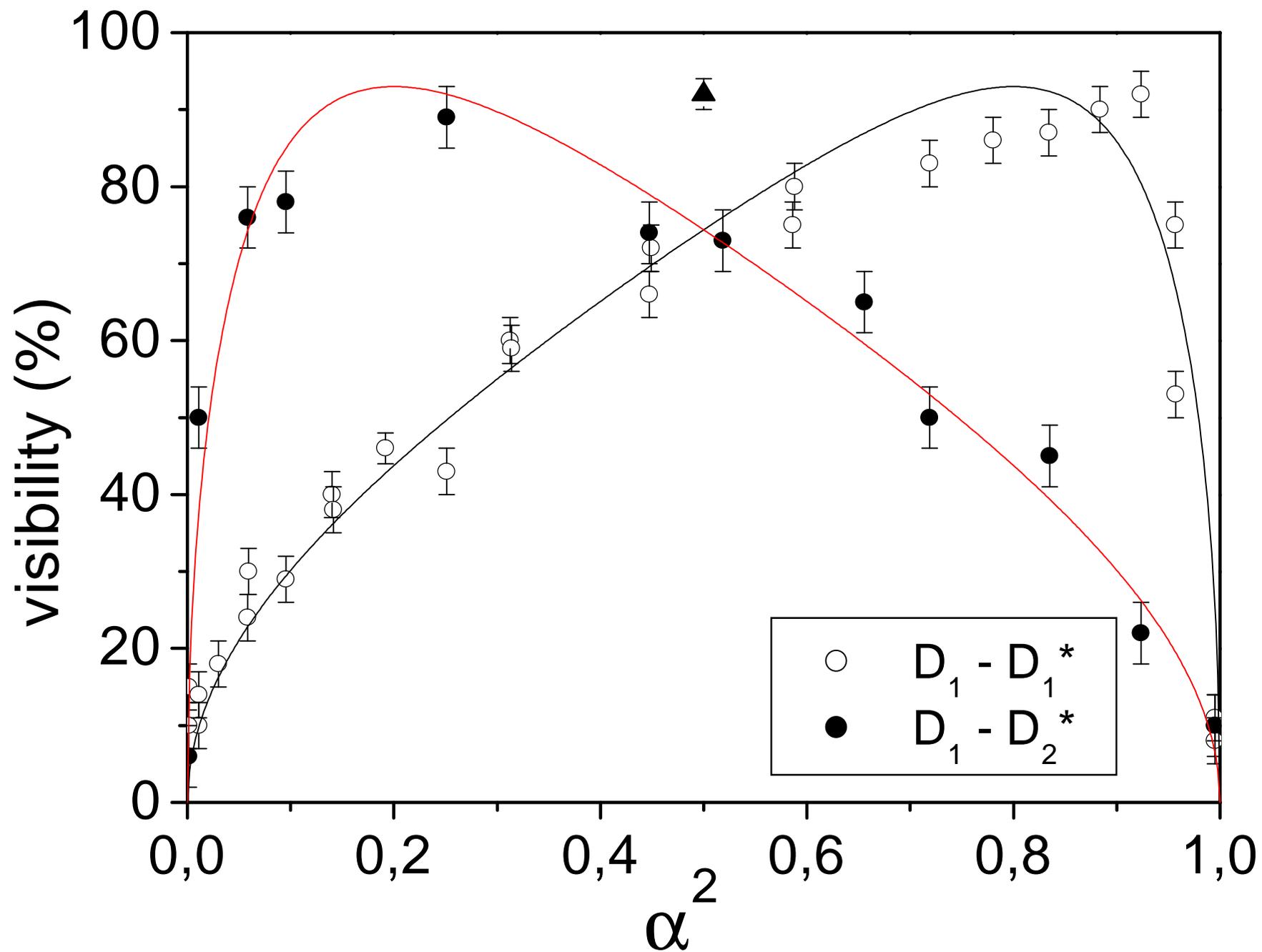
\end{document}